\documentclass[a4paper,10pt]{article}
\usepackage[normalem]{ulem} 
\usepackage[makeroom]{cancel} 
\usepackage{amsmath,amssymb,mathtools}  
\usepackage{soul}                       
\usepackage[usenames,dvipsnames]{xcolor}
\usepackage[raggedright]{titlesec}
\usepackage{scrextend}
\deffootnote{2em}{0em}{\thefootnotemark\quad}
\usepackage{multicol}
\usepackage[left=0.75in,right=0.75in,top=1.0in,bottom=1.25in]{geometry}
\usepackage{hyperref}
\usepackage{float}
\usepackage{xcolor}
\usepackage{booktabs}
\usepackage{caption}
\expandafter\let\csname equation*\endcsname\relax
\expandafter\let\csname endequation*\endcsname\relax

\hypersetup{
  colorlinks   = true, 
  urlcolor     = blue, 
  linkcolor    = blue, 
  citecolor   = blue 
}

\definecolor{seablue}{rgb}{0.0, 0.4784, 0.6471}
\definecolor{deepskyblue}{rgb}{0.0, 0.75, 1.0}
\newcommand{\dd}{\text{d}}

\newcommand{\dx}{\text{d}x}

\setstcolor{red}    
\setulcolor{red}    
\allowdisplaybreaks 

\let\oldabstract\abstract
\let\oldendabstract\endabstract
\makeatletter
\renewenvironment{abstract}
{%
               {\list{}{\addtolength{\leftmargin}{4em} 
                        \listparindent 1.5em%
                        \itemindent    \listparindent%
                        \rightmargin   \leftmargin%
                        \parsep        \z@ \@plus\p@}%
                \item\relax}%
               {\endlist}%
\oldabstract}
{\oldendabstract}
\makeatother

\title{\LARGE\textbf{\textsf{On the KG-constrained Bekenstein's disformal transformation\\ of the Einstein-Hilbert action}}}

\author{\normalsize Allan L. Alinea\footnote{main and corresponding author: alalinea@up.edu.ph} \;and Joshwa DJ. Ordo\~nez}

\date{}

\begin{document}

\maketitle
\vspace{-2.25em}
\noindent
\begin{center}
{\small Astrophysics, Particle Physics, and Nuclear Physics Research Cluster\\Institute of Physics, \\University of the Philippines Los Ba\~nos\\4031 College, Los Ba\~nos, Laguna, Philippines}
\end{center}

\bigskip
\begin{abstract}
\noindent
Motivated by an inclination for symmetry and possible extension of the General Theory of Relativity within the framework of Scalar Theory, we investigate the Bekenstein's disformal transformation of the Einstein-Hilbert action. Owing to the complicated combinations of second order metric derivatives encoded in the Ricci scalar of the action, such a transformation yields an unwieldy expression. To {``tame''} the transformed action, we exploit the {Klein-Gordon (KG)} conformal-disformal constraint previously discovered in the study of the invariance of the massless Klein-Gordon equation under disformal transformation. The result upon its application is a surprisingly much more concise and simple action in four spacetime dimensions containing three out of four sub-Lagrangians in the Horndeski action, and three beyond-Horndeski terms. The latter group of terms may be attributed to the kinetic-{term} dependence of the conformal and disformal factors in the Bekenstein's disformal transformation. Going down to three dimensions, we find a relatively simpler resulting action but the signature of the three {``extraneous''} terms remains. Lastly, in two dimensions, we find an invariant action consistent with its topological nature in these dimensions.

\par
\vspace{0.75em}
\noindent
{\textbf{keywords:} \textit{disformal transformation, symmetry, Einstein-Hilbert action, Klein-Gordon\\ \phantom{dominatedxx}equation, Horndeski action, scalar-tensor theory}}	
\end{abstract}

\bigskip
\begin{multicols}{2}
\section{Introduction}
\label{secIntro}
\noindent
Our search for symmetries was once confined to tangible objects and physical phenomena. General Relativity expanded our search to include the fabric of spacetime itself. Aside from general covariance, it suggests that we also ought to look for symmetries of covariant field equations, actions, etc. under ``warpings'' of spacetime---under metric transformations. Symmetries under the well-established \emph{conformal transformation} were the first to be discovered. {``Canonical''} entities of General Relativity such as the Weyl tensor, and divergence theorem of the traceless energy-momentum tensor remain invariant under conformal transformations. \cite{Wald:1984rg} The same is true for Maxwell's equations, \cite{Cunningham:1910,Bateman:1910} the Klein-Gordon equation in two dimensions \cite{Madsen:1993ke}, and primordial cosmological perturbations \cite{Fakir:1990eg, Makino:1991sg, Kubota:2011re}---the seeds that gave rise to the large-scale structures in the Universe that we observe today.

In an attempt to couple a scalar field to gravity, many have proposed gravitational theories with two metrics---one for gravity and one for matter dynamics---that often turn out to be conformally related \cite{Fujii:2003pa, Capozziello:1998dq,Maeda:1988ab}. This conformal mapping allows one to cast scalar-coupled gravitational actions, residing in what is called the \emph{Jordan frame}, into the more convenient \emph{Einstein frame} where the scalar field couples only to the matter action and the gravitational action is 
\begin{align}
    S
    =
    \frac{1}{2}\int \dd^n x\,
    \sqrt{-g}\, R,
\end{align}
which is the \emph{Einstein-Hilbert action} of General Relativity \cite{Fujii:2003pa, Faraoni:2010pgm}. In light of this conformal mapping and its associated symmetries, the conformal transformation is a powerful tool to probe new gravitational theories with scalar coupling.\cite{Gionti:2017ffe,Nozari:2009ds} The early investigation of these ideas marked the birth of \emph{Scalar-Tensor theory} \cite{Nordstrom:1913dga,Brans:1961sx,Dirac:1938mt,Kaluza:1921tu,Klein:1926tv}.

A question naturally follows: Is the conformal transformation the most general metric transformation that exhibits symmetries and connects gravitational metrics to viable matter metrics? To answer this question, Bekenstein imposed physical constraints on a Finslerian metric, which may be regarded as the most general incarnation of the metric in his time. In the end, his equation for the Finslerian metric is reduced to a relation between two Riemannian metrics \cite{Bekenstein:1992pj} and is now recognized as the \emph{{Bekenstein's} disformal transformation}:
\begin{align}
    \label{bekdisftransfo}
    g_{\mu\nu}
    \,\rightarrow\,
    \widehat g_{\mu\nu}
    =
    A(\phi,X)g_{\mu\nu}
    +
    B(\phi,X) \phi_{;\mu}\phi_{;\nu},
\end{align}
where $X$ is the kinetic term defined by $X\equiv-g_{\mu\nu}\phi^{;\mu}\phi^{;\nu}/2$. Notice that the disformal transformation reduces to the conformal transformation when $B=0$. Unlike the conformal transformation, which scales distances isotropically, the disformal transformation preferentially scales distances along the gradient of the scalar field. As such, the angle between two tangent vectors at a single point in a spacetime manifold is in general, not preserved under disformal transformation.

There is no \emph{a priori} reason why the Universe would prefer symmetries and mappings under \eqref{bekdisftransfo} with $B=0$; {that is, conformal transformation alone.} It is then reasonable to hypothesize that the symmetries under the conformal transformation are special cases of more general symmetries under the Bekenstein disformal transformation. Many have investigated this idea soon after Bekenstein published his work. It was shown that Maxwell's equations \cite{Goulart:2013laa}, the massless Klein-Gordon equation (under the KG conformal-disformal constraints) \cite{Falciano:2011rf,Alinea:2022ygr}, the Dirac Equation (under the Inomata condition) \cite{Bittencourt:2015ypa}, and the gauge-invariant primordial cosmological perturbations \cite{Tsujikawa:2014uza,Domenech:2015hka,Motohashi:2015pra,Alinea:2020laa} are invariant under the Bekenstein's disformal transformation and its generalizations. 

In the context of Scalar-Tensor theory, we expect the disformal transformation to map gravitational metrics to a more general class of matter metrics. Viable actions associated with matter metrics are constrained by Ostrogradski instabilities that arise from higher than second-order terms in the equations of motion \cite{Woodard:2015zca,Ostrogradsky:1850fid}. The most general second-order Scalar-Tensor action that leads to second-order equations motion in four dimensions is the \emph{Horndeski action} \cite{Horndeski:1974wa}. It has been shown \cite{Bettoni:2013diz} that the Horndeski action is {\emph{form-invariant}} under {the \emph{special disformal transformation} given by} 
\begin{align}
    \label{specdisftransfo}
    g_{\mu\nu}
    \,\rightarrow\,
    \widehat g_{\mu\nu}
    =
    A(\phi)g_{\mu\nu}
    +
    B(\phi) \phi_{;\mu}\phi_{;\nu}.
\end{align}
{Needless to say, this is a special case of the Bekenstein's disformal transformation where the conformal and disformal factors have no dependence on the kinetic term, $ X $}. This transformation allows us to map gravitational and matter metrics encapsulated by the Horndeski action.

A special case of the Horndeski action is the Einstein-Hilbert action. The Horndeski action contains derivative {``counter-terms''} that play a role in making it form-invariant under the special disformal transformation \cite{Bettoni:2013diz}, and it is not obvious { if} the Einstein-Hilbert action {would remain} invariant or would map to a Horndeski action despite the absence of these terms. However, we have shown in a previous study that the Einstein-Hilbert action {completely} maps to a Horndeski action under the special disformal transformation \cite{Alinea:2020sei}; {in particular, the resulting Lagrangians belong to three of the four sub-Lagrangians of the Horndeski action.} This result allows us to cast special cases of the Horndeski action that may be of interest for a metric where the action takes a convenient form. In other words, there exists an analogous Jordan frame to Einstein frame mapping under disformal transformations. Conversely, we may use the special disformal transformation of the Einstein-Hilbert action to probe guaranteed-viable (no Ostrogradsky instability) Scalar-Tensor actions.

The significance of this mapping motivates us to investigate how the Einstein-Hilbert action transforms under more general metric transformations. For instance, we may consider the \emph{general disformal transformation}, which is the generalization of the Bekenstein disformal transformation that includes second-order derivatives in the scalar field. The additional degrees of freedom from this transformation inevitably lead to non-Horndeski terms in the transformed Einstein-Hilbert action. These extraneous terms may be removed by imposing constraints on the functionals of the action (see for instance, Refs. \cite{Alinea:2022ygr,Takahashi:2021ttd}), or they may be assimilated by identifying them with terms that belong to healthy \emph{beyond-Horndeski actions} \cite{Gleyzes:2014dya,Gleyzes:2014qga,BenAchour:2016cay,Langlois:2015cwa,Kobayashi:2019hrl,Takahashi:2022mew,Domenech:2015tca,Takahashi:2017zgr}. We investigated the general (but constrained) disformal transformation of the Einstein-Hilbert action in Ref. \cite{Alinea:2024gjn}.

The natural continuation of Ref. \cite{Alinea:2020sei} {involving the special disformal transformation of the Einstein-Hilbert action,} is to consider the Bekenstein disformal transformation of the Einstein-Hilbert action. {However, as could be predicted from the study in Ref. \cite{Bettoni:2013diz}, such a transformation could lead to beyond-Horndeski terms and as initially thought, Ostrogradski instabilities; although the latter, can be addressed within the framework of degenerate beyond-Horndeski theories\cite{Gleyzes:2014dya,Gleyzes:2014qga,BenAchour:2016cay,Langlois:2015cwa,Kobayashi:2019hrl,Takahashi:2022mew,Domenech:2015tca,Takahashi:2017zgr}}, we anticipate that due to the second-order metric derivative terms encoded in the Ricci scalar, the resulting transformed Einstein-Hilbert action would be lengthy and complicated.\footnote{See also Ref. \cite{BenAchour:2016cay} for a related work on the Bekenstein's disformal transformation involving a more general scalar-tensor action, degeneracy, and classifications.} The number of beyond-Horndeski terms would far exceed those that could be mapped to the Horndeski action. Having said this, driven by curiosity, it would be interesting to know if there are constraints or conditions that we may impose on the conformal and disformal factors to simplify the resulting action. The ideal aim is to completely remove all beyond-Horndeski terms. Nonetheless, a  simplification with few remaining beyond-Horndeski terms would be a significant relief on our side. Besides, the few remaining beyond-Horndeski terms may find utility in current and future theories that could shed more light about gravity.

With regards to the set of constraints that we may impose in simplifying the transformed action, our guiding idea is that this should at least be respecting another family member belonging under the same wing of the Horndeski action. At a glance, the Klein-Gordon and Einstein-Hilbert actions may be  mathematically and physically different. However, they are ``siblings'' under the family of Horndeski action. As such, given that the Klein-Gordon equation is invariant under the disformal transformation  with the Klein-Gordon conformal-disformal constraint \cite{Falciano:2011rf,Alinea:2022ygr,Alinea:2024gjn}, we find it interesting to apply the same constraint for the disformal transformation of the Einstein-Hilbert action. Anticipating a possible shared symmetry, we hypothesize that the transformed Einstein-Hilbert action under such constraint, if not invariant or a proper subset of the Horndeski action, would contain only a few beyond-Horndeski terms. As we shall see below, the complete invariance is attained in $ n = 2 $ dimensions similar to that for conformal transformation and consistent with the topological nature of this action in these dimensions. On the other hand, we are successful in significantly ``taming'' the transformed Einstein-Hilbert action in $ n = 3 $ and $ n = 4 $ dimensions.


In this study, we derive the Bekenstein disformal transformation of the Einstein-Hilbert action subject to the conformal-disformal constraint that leads to the disformal invariance of the massless Klein-Gordon equation. Our paper is organized as follows. In Sec. \ref{seckgdisf} we review the Bekenstein disformal transformation of the massless Klein-Gordon equation and the conformal-disformal constraint that leads to its invariance. Afterwards, we use the same constrained transformation on the Ricci tensor in Sec. \ref{secdtrictens}. In Sec. \ref{secdtehact}, we calculate the transformed Ricci scalar and the transformed Einstein-Hilber action. We show that the transformed action consists of Horndeski terms and three beyond-Horndeski terms. We also show that the coefficients of the beyond-Horndeski terms simplify in $n=3$ dimensions, and that the Einstein-Hilbert action is \emph{invariant} in $n=2$ dimensions. Lastly, we give our concluding remarks and future prospects in Sec. \ref{seConclude}. 
\section{Bekenstein's disformal transformation and the massless Klein Gordon equation}
\label{seckgdisf}
\noindent
The Klein-Gordon equation for a massless scalar field $\phi$ states $\square \phi = 0$, where the d'Alembertian is defined as $\square \equiv g^{\mu\nu}\nabla_\mu\nabla_\nu$, with $\nabla_\mu$ being the covariant derivative operator with respect to the spacetime coordinate $x^\mu$. In this section, we provide a relatively concise derivation of its invariance under the Bekenstein's disformal transformation, $\widehat g_{\mu\nu} = A(\phi, X) g_{\mu\nu} + B(\phi,X)\phi_{;\mu} \phi_{;\nu},$ subject to the conformal-disformal constraint given by \cite{Falciano:2011rf,Alinea:2022ygr,Alinea:2024gjn}
\begin{align}
    \label{confdisfcon}
    B(\phi,X)
    =
    \frac{A - b^2 A^{n-1}}{2X}
    \quad
    (b = \text{const.}),
\end{align}
where $n$ is the number of spacetime dimensions. This constraint relates the disformal factor to the conformal factor, effectively making $A$ a ``free'' transformation factor which determines\footnote{To be clear, as elucidated in Ref. \cite{Alinea:2022ygr} and consistent with the discussion in Refs. \cite{Falciano:2011rf, Alinea:2024gjn} this may not be interpreted as an additional equation for $\phi$. Such an interpretation could lead to an over-determined $\phi$.} $B$.  

Under the Bekenstein's disformal transformation, the d'Alembertian of $\phi$ changes as $ \square\phi \rightarrow \widehat \square \phi $, where $\widehat \square \phi = \widehat g^{\mu\nu}\widehat \nabla_\mu \widehat \nabla_\nu \phi$, with the inverse disformal metric satisfying the contraction equation given by, $\widehat g^{\mu\alpha} \widehat g_{\alpha\nu} = \delta^\mu_\nu$. Using the Sherman–Morrison formula, one easily finds the inverse metric as
\begin{align}
    \label{hatgmunu}
	\widehat g^{\mu \nu }
	&=
	\frac{g^{\mu \nu }}{A} 
	-
	\frac{B\phi ^{;\mu }\phi ^{;\nu }}{A(A - 2BX)}.
\end{align}
On the other hand, the hatted second order covariant derivative of the scalar field may be expressed as $	\widehat \nabla _\mu \widehat \nabla _{\nu }\phi = \phi _{;\nu,\mu } - \widehat \Gamma ^\alpha _{\mu \nu }\phi _{;\alpha }$, where the hatted Christoffel symbol follows the same form as the original (unhatted) one; specifically,
$\widehat \Gamma^\alpha_{\mu\nu} = \tfrac{1}{2}\widehat g^{\alpha\beta}(\widehat g_{\beta\mu,\nu} + \widehat g_{\nu\beta,\mu} - \widehat g_{\mu\nu,\beta})$. Interestingly, it decomposes into the original Christoffel symbol and disformal contribution terms; symbolically, $\widehat \Gamma ^\alpha _{\mu \nu } = \Gamma ^\alpha _{\mu \nu }	+ C^\alpha _{\mu \nu },$ where the second term captures the disformal contributions involving the conformal and disformal factors, scalar field, and their derivatives. In particular, this term is given by 
\begin{align}
    \label{Calphamunu}
	C^\alpha _{\mu \nu }	
	&=
	\frac{
		A_{;\nu }\delta ^\alpha _\mu 
		+
		A_{;\mu }\delta ^\alpha _\nu 
		-
		A^{;\alpha }g_{\mu \nu }	
	}{2A }
	\nonumber
	\\[0.5em]
	&\quad 
	-\,
	\frac{
		B\phi^{;\alpha}(
			A _{;\nu }\phi _{;\mu} 
			+
			A _{;\mu }\phi _{;\nu} 
			-
			A _{;\beta }\phi^{;\beta} g_{\mu \nu }
		)
	}{2A (A - 2BX)} 
	\nonumber
	\\[0.5em]
	&\quad
	+\,
	\left[
		\frac{g^{\alpha \beta }}{A}
		-
		\frac{B\phi^{;\alpha} \phi ^{;\beta} }{A(A - 2BX )} 
	\right]
	\left(
		\tfrac{1}{2} \phi _{;\beta }\phi _{(;\mu }B_{;\nu )}
		-
		\tfrac{1}{2} B_{;\beta }\phi _{;\mu }\phi _{;\nu }
	\right)
	\nonumber
	\\[0.5em]
	&\quad
	+\,
	\frac{
		\tfrac{1}{2} \phi ^{;\alpha }B_{(;\mu }\phi _{;\nu )}
		+
		B\phi ^{;\alpha }\phi _{;\mu \nu }
	}{A - 2BX},
\end{align}
where the parenthesised covariant derivative indices indicate symmetric combination; e.g., $\phi _{(;\mu }B_{;\nu )} = \tfrac{1}{2}(\phi _{;\mu }B_{;\nu} + \phi _{;\nu }B_{;\mu})$. Observe that $C^\alpha_{\mu\nu}$ is symmetric with respect to the lower indices, $(\mu,\nu)$. As such, in torsion-free spacetime, $\widehat \Gamma^\alpha_{\mu\nu}$ inherits the symmetry property with respect to its lower indices from the original Christoffel symbol. We further note that $C^\alpha_{\mu\nu}$, being a difference of two Christoffel symbols, is tensorial in nature.

Using the equations above for $\widehat \Gamma^\alpha_{\mu\nu}$ and $\widehat g^{\mu\nu}$ in the equation for the hatted double covariant derivative of $\phi$ yields $\widehat \nabla_\mu \widehat \nabla_\nu \phi = \nabla_\mu \nabla_\nu \phi - C^\alpha _{\mu\nu} \phi_{;\alpha}.$ In other words, the disformally transformed second order covariant derivative of the scalar field decomposes into the original second order covariant derivative of $\phi$ and disformal contributions terms. Note however, that unlike the decomposition of $\widehat \Gamma^\alpha_{\mu\nu}$ where $\Gamma^\alpha_{\mu\nu}$ is nowhere to be found in second term involving the disformal contributions, $\nabla_\mu \nabla_\nu \phi$ is present in $C^\alpha _{\mu\nu} \phi_{;\alpha}$. The last term in the equation above for $C^\alpha _{\mu\nu}$ given by (\ref{Calphamunu}) spoils a ``clean-cut'' decomposition of $\widehat \nabla_\mu \widehat \nabla_\nu \phi$ into $\nabla_\mu \nabla_\nu \phi$  and terms explicitly involving only up to first order derivative of the scalar field.

Having said this, the second order term in $C^\alpha _{\mu\nu}$, namely, $B\phi^{;\alpha}\phi_{;\mu\nu}/(A - 2BX)$, simply combines with $\phi_{;\mu\nu}$ in the first term on the right hand side of $\widehat \nabla_\mu \widehat \nabla_\nu \phi = \nabla_\mu \nabla_\nu \phi - C^\alpha _{\mu\nu} \phi_{;\alpha},$ upon calculation of $\widehat \square \phi$ to form $\square \phi$ and a term involving $\phi^{;\alpha}X_{;\alpha}$. The full result of the calculation for $\widehat \square \phi = \widehat g^{\mu\nu} \widehat \nabla_\mu \widehat \nabla_\nu \phi$ is given by 
\begin{align}
	\label{dispsqphi}
	\widehat\square\phi 
	&=
	\frac{\square \phi }{A - 2BX} 
	+
	\frac{B\phi ^{;\alpha }X_{;\alpha }}{(A - 2BX)^2} 
	+
	\frac{X\phi ^{;\alpha }B_{;\alpha }}{(A - 2BX)^2} 
    \nonumber
    \\[0.5em]
	&\quad
    +\,
	\frac{n(A - 2BX) - 2A + 2BX}{2A(A - 2BX)^2}
	\phi ^{;\alpha }A_{;\alpha }.
\end{align}
Given the length and relatively complicated nature of the expression for $C^\alpha_{\mu\nu}$ as given by (\ref{Calphamunu}), this result is remarkably simple. 

In the conformal limit, $B  = 0$,  the equation above reduces to
\begin{align}
	\widehat\square\phi 
	&\rightarrow
	\frac{\square \phi }{A} 
	+
	\frac{n-2}{2A^2}
	\phi ^{;\alpha }A_{;\alpha }.
\end{align}
This tells us that \textit{in general}, the massless Klein-Gordon equation, is not invariant under the conformal transformation. On the other hand, under the \textit{special} disformal transformation, where both $A$ and $B$ depends on $\phi$ alone, we have
\begin{align}
	\widehat \square \phi 
	&\rightarrow	
    \frac{\square \phi }{A - 2BX} 
    -
    \frac{nA_{,\phi}X}{A(A - 2BX)} 
    +
    \frac{B\phi ^{;\alpha }X_{;\alpha }}{(A - 2BX)^2}    
    \nonumber
    \\[0.5em]
    &\qquad
    -\,
    \frac{2X[(AB)_{,\phi}X - AA_{,\phi}]}{A(A - 2BX)^2}.
\end{align} 
Because of the existence of the kinetic term $X$ in the special disformal limiting form above, there is no non-trivial equation relating $B$ to $A$ that can make $\widehat \square \phi \sim \square \phi$. In other words, the massless Klein-Gordon equation is in general, variant under the conformal and special disformal transformations. 

When the special disformal transformation is generalised to the Bekenstein's disformal transformation, we can cancel all the terms on the right hand side of (\ref{dispsqphi}) beyond the first one involving $\square\phi$. Indeed, performing functional separation of the disformal factor into $B = f(A)\,h(X),$ where $f$ and $h$ are functionals of $A(\phi,X)$ and $X$, respectively, one finds
\begin{align}
	\phi ^{;\alpha }B_{;\alpha } 
	=
	f h_{,X} \phi ^{;\alpha } X_{;\alpha }	
	+
    f_{,A} h \phi ^{;\alpha } A_{;\alpha },
\end{align}
for the third term on the right hand side of (\ref{dispsqphi}) to correspondingly cancel the other two terms involving $\phi^{;\alpha}X_{;\alpha}$ and $\phi^{;\alpha}A_{;\alpha}$.
This leads to a system of functional differential equations given by
\begin{align}
    h + h_{,X}X &= 0,
    \nonumber
    \\[0.5em]
    2hAXf_A - 2(n-1)hXf + (n-2)A &= 0.
\end{align}
The solutions of this system of differential equations can be written as
\begin{align}
    h &= \frac{1}{X}
    \quad\text{and}\quad
    f = \frac{A - b^2 A^{n-1}}{2},
\end{align}
which combine to form the conformal-disformal constraint given by 
(\ref{confdisfcon}) above. $\blacksquare$

In perspective, under the Bekenstein's disformal transformation subject to the conformal-disformal constraint (\ref{confdisfcon}), the transformed Klein-Gordon equation is related to the original one as
\begin{align}
    \square \phi + m^2\phi = 0
    \quad\rightarrow\quad
    \widehat \square \phi + \frac{m^2}{b^2A^{n-1}}\phi = 0.
\end{align}
We find the mass $m^2$ is scaled by a factor $1/b^2A^{n-1}$ upon transformation. Consequently, when $m$ vanishes, the massless Klein-Gordon equation is invariant under the constrained Bekenstein's disformal transformation.

\section{Transformation of the Ricci tensor}
\label{secdtrictens}

\noindent 
The hatted Riemann curvature tensor follows the same form as the original one. In terms of the hatted Christoffel symbols and derivatives thereof, we may write
\begin{align}
	\widehat R^\alpha _{\mu \beta \nu }
	&=
	-\widehat \Gamma ^\alpha _{\mu \beta ,\nu }
	+
	\widehat \Gamma ^\alpha _{\mu \nu ,\beta }
	-
	\widehat \Gamma ^\rho _{\mu \beta }\widehat \Gamma ^\alpha _{\rho \nu }
	+
	\widehat \Gamma ^\rho _{\mu \nu  }\widehat \Gamma ^\alpha _{\rho \beta  }	
\end{align}
The hatted (transformed) Ricci tensor---our object in this section---follows from this expression through the contraction of the first and third indices. Because the hatted Christoffel symbol decomposes as $\widehat \Gamma^\alpha_{\mu\nu} = \Gamma^\alpha_{\mu\nu} + C^\alpha_{\mu\nu}$, the hatted Ricci tensor correspondingly fragments as
\begin{align}
    \label{hatRmunu}
	\widehat R_{\mu \nu }
	&=
	R_{\mu \nu }
	-
	C ^\alpha _{\mu\alpha;\nu }
	+
	C ^\alpha _{\mu\nu;\alpha  }
	-
	C ^\rho _{\mu\alpha} C^\alpha _{\rho \nu }
	+
	C ^\rho _{\mu\nu} C^\alpha _{\rho \alpha}.	
\end{align}
Given the form of $C^\alpha_{\mu\nu}$ from (\ref{Calphamunu}), we find the four disformal contribution terms in the equation above for $\widehat R_{\mu\nu}$, symmetric with respect to the indices $(\mu,\nu)$. It follows that $\widehat R_{\mu\nu}$ inherits the symmetry property of $R_{\mu\nu}$ with respect to its two indices.

Equation (\ref{Calphamunu}) tells us a rather lengthy expression for $C^\alpha_{\mu\nu}$. This, in turn, suggests a complicated resulting equation for $\widehat R_{\mu\nu}$ upon substitution from (\ref{Calphamunu}) in (\ref{hatRmunu}). Indeed, after a long but straightforward calculation, we find
\begin{align}
	\label{hatRmnRmn}
	&\widehat R_{\mu \nu } - R_{\mu \nu }
	\nonumber
	\\[0.5em]
	&=
	\frac{B^2 X_{;\mu} X_{;\nu}}{(A - 2BX)^2} 
	+ 
	\frac{A {B_{;\mu }} X_{;\nu}}{2 (A - 2BX)^2} 
	- 
	\frac{B A_{;\mu} X_{;\nu}}{2 (A - 2BX)^2} 
	\nonumber
	\\[0.5em]
	&\quad
	-\,
	\frac{B^{;\alpha} A \phi_{;\alpha\mu}  \phi_{;\nu}}{2 A (A - 2BX)} 
	+
	\frac{B X \phi^{;\alpha} B_{;\alpha} \phi^{;\beta} {B_{;\beta}} \phi_{;\mu} \phi_{;\nu}}{2 A (A - 2BX)^2} 
	\nonumber
	\\[0.5em]
	&\quad 
	+\,
	\frac{B \phi^{;\alpha} B_{;\alpha\beta} \phi^{;\beta} \phi_{;\mu} \phi_{;\nu}}{2 A (A - 2BX)} 
	- 
	\frac{(\square B) \phi_{;\mu} \phi_{;\nu}}{2 A}
	\nonumber
	\\[0.5em]
	&\quad
	+\, 
	\frac{B[n(A - 2BX) - 2(2A - 3BX)] 
		\phi^{;\alpha} A_{;\alpha} \phi ^{;\beta }B_{;\beta }
		\phi_{;\mu} \phi_{;\nu}}{4 A^2 (A - 2BX)^2} 
	\nonumber
	\\[0.5em]
	&\quad
    +\, 
	\frac{B\, (\square\phi) \phi^{;\alpha} B_{;\alpha} \phi_{;\mu} \phi_{;\nu}}{2 A (A - 2BX)}
	+
	\frac{B^2 \phi^{;\alpha} B_{;\alpha} \phi ^{;\beta }X_{;\beta} \phi_{;\mu} \phi_{;\nu}}{2 A (A - 2BX)^2}  
	\nonumber
	\\[0.5em]
	&\quad	
	-\,
	\frac{X B^{;\alpha} B_{;\alpha} \phi_{;\mu} \phi_{;\nu}}{2 A (A - 2BX)} 
	-
	\frac{B A^{;\alpha }A_{;\alpha }\phi_{;\mu} \phi_{;\nu}}{2 A^2 (A - 2BX)}  
	\nonumber
	\\[0.5em]
	&\quad
	-\,
	\frac{[n(A - 2BX) - 6(A - BX)] A^{;\alpha} B_{;\alpha} \phi_{;\mu} \phi_{;\nu}}{4 A^2 (A - 2BX)}
	\nonumber
	\\[0.5em]
	&\quad 
	+
	\frac{X \phi^{;\alpha} B_{;\alpha} B_{;\mu} \phi_{;\nu}}{2 (A - 2BX)^2}
	+
	\frac{(\square\phi) B_{;\mu} \phi_{;\nu}}{2 (A - 2BX)}  
    \nonumber
	\\[0.5em]
	&\quad	 
	+\,
	\frac{[n(A - 2BX) - 2(2A - 3BX)]
		\phi ^{;\alpha }A_{;\alpha } B_{;\mu} \phi_{;\nu}}{4 A (A - 2BX)^2}
	\nonumber
	\\[0.5em]
	&\quad
	+\, 
	\frac{B\, \phi ^{;\alpha }X_{;\alpha} B_{;\mu} \phi_{;\nu}}{2 (A - 2BX)^2} 
	-
	\frac{(A - B X)  \phi^{;\alpha} B_{;\alpha} A_{;\mu} \phi_{;\nu}}{2 A (A - 2BX)^2}
	\nonumber
	\\[0.5em]
	&\quad	
	-\,
	\frac{B[n(A - 2BX) - 2(3A - 5BX)]  \phi ^{;\alpha }A_{;\alpha }
		A_{;\mu} \phi_{;\nu}}{4 A^2 (A - 2BX)^2} 
	\nonumber
	\\[0.5em]
	&\quad  
	-\,
	\frac{B\, (\square\phi) A_{;\mu} \phi_{;\nu}}{2 A (A - 2BX)} 
	- 
	\frac{B^2 \phi ^{;\alpha }X_{;\alpha }A_{;\mu} \phi_{;\nu}}{2 A (A - 2BX)^2}
	\nonumber
	\\[0.5em]
	&\quad	
	+\,
	\frac{B A^{;\alpha} \phi_{;\alpha\mu} \phi_{;\nu}}{2 A (A - 2BX)} 
	+ 
	\frac{\phi^{;\alpha} B_{;\alpha\mu} \phi_{;\nu}}{2 (A - 2BX)} 
	- 
	\frac{B \phi^{;\alpha} A_{;\alpha\mu} \phi_{;\nu}}{2 A (A - 2BX)} 
	\nonumber
	\\[0.5em]
	&\quad 
	+\,
	\frac{A X_{;\mu} B_{;\nu}}{2 (A - 2BX)^2} 
	+ 
	\frac{X \phi^{;\alpha} B_{;\alpha} \phi_{;\mu} B_{;\nu}}{2 (A - 2BX)^2}
	\nonumber
	\\[0.5em]
	&\quad
	+\,
	\frac{[n(A - 2BX) - 2(2A - 3BX)]
		\phi ^{;\alpha }A_{;\alpha } \phi_{;\mu} B_{;\nu}}{4 A (A - 2BX)^2}
	\nonumber
	\\[0.5em]
	&\quad	  
	+\, 
	\frac{(\square\phi) \phi_{;\mu} B_{;\nu}}{2 (A - 2BX)} 
	+ 
	\frac{B\, \phi ^{;\alpha }X_{;\alpha }\phi_{;\mu} B_{;\nu}}{2 (A - 2BX)^2}
	+
	\frac{X^2 B_{;\mu} B_{;\nu}}{(A - 2BX)^2}  
	\nonumber
	\\[0.5em]
	&\quad
	-\,
	\frac{X\,(A - BX) A_{;\mu} B_{;\nu}}{A(A - 2BX)^2} 
	- 
	\frac{B X_{;\mu} A_{;\nu}}{2 (A - 2BX)^2} 
	\nonumber
	\\[0.5em]
	&\quad		 
	-\,
	\frac{(A - 6 B X) \phi_{;\alpha} B_{;\alpha} \phi_{;\mu} A_{;\nu}}{8 A^2 (A - 2BX)}
	-
	\frac{X\, (A - BX)  B_{;\mu} A_{;\nu}}{A (A - 2BX)^2} 
	\nonumber
	\\[0.5em]
	&\quad		
	-\,
	\frac{[4BX(A - 3BX) + 3 A^2] \phi^{;\alpha} B_{;\alpha} \phi_{;\mu} A_{;\nu}}{8 A^2 (A - 2BX)^2}
 	+ 
	\frac{B \phi^{;\alpha} \phi_{;\mu\nu\alpha}}{A - 2BX}
	\nonumber
	\\[0.5em]
	&\quad		 
	-
	\frac{B[n(A - 2BX) - 2(3A - 5BX)]\phi ^{;\alpha }A_{;\alpha }
		\phi_{;\mu} A_{;\nu}}{4 A^2 (A - 2BX)^2}
	\nonumber
	\\[0.5em]
	&\quad		
	-\,
	\frac{B\, (\square\phi) \phi_{;\mu} A_{;\nu}}{2 A (A - 2BX)} 
	- 
	\frac{B^2 \phi ^{;\alpha }X_{;\alpha }\phi_{;\mu} A_{;\nu}}{2 A (A - 2BX)^2} 
    + 
    A_{;\mu} A_{;\nu}
	\nonumber
	\\[0.5em]
	&\quad	
	\times\,
	\frac{[12(n-3)(BX)^2 - 12 A B X n + 3(n-2)A^2 + 32 A B X]  }{4 A^2 (A - 2BX)^2}
	\nonumber
	\\[0.5em]
	&\quad	
	+\,
	\frac{B \phi^{;\alpha} A_{;\alpha\beta} \phi^{;\beta} g_{\mu\nu}}{2 A (A - 2BX)} 
	+
	\frac{(A - BX)\phi ^{;\alpha }A_{;\alpha } \phi^{;\beta } B_{;\beta } g_{\mu\nu}}{2 A (A - 2BX)^2} 
	\nonumber
	\\[0.5em]
	&\quad 
	+\,
	\frac{X A^{;\alpha} B_{;\alpha} g_{\mu\nu}}{2 A (A - 2BX)}
	+ 
	\frac{B\,\phi ^{;\alpha }A_{;\alpha }(\square\phi) g_{\mu\nu}}
		{2 A (A - 2BX)}  
	- 
	\frac{(\square A) g_{\mu\nu}}{2 A}  
	\nonumber
	\\[0.5em]
	&\quad
	+\,
	\frac{B[n(A - 2BX) - 3(2A - 5BX)]
		\phi ^{;\alpha }A_{;\alpha }\phi ^{;\beta }A_{;\beta }g_{\mu\nu}}{4 A^2 (A - 2BX)^2}
	\nonumber
	\\[0.5em]
	&\quad		
	-\,
	\frac{[n(A - 2BX) - 2(2A - 5BX)]A^{;\alpha }A_{;\alpha }
		g_{\mu\nu}}{4 A^2 (A - 2BX)} 
	\nonumber
	\\[0.5em]
	&\quad	
	+\,
	\frac{B^2 \phi ^{;\alpha }A_{;\alpha }\phi ^{;\beta }X_{;\beta } g_{\mu\nu}}
		{2 A (A - 2BX)^2} 
	+ 
	\frac{B X_{;\mu\nu}}{(A - 2BX)} 
	+ 
	\frac{B\, (\square\phi) \phi_{;\mu\nu}}{(A - 2BX)}  
	\nonumber
	\\[0.5em]
	&\quad	 
	+\,
	\frac{(A - BX)  \phi^{;\alpha} B_{;\alpha} \phi_{;\mu\nu}}{(A - 2BX)^2}
	+ 
	\frac{B^2 \phi ^{;\alpha }X_{;\alpha }\phi_{;\mu\nu}}{(A - 2BX)^2} 
	\nonumber
	\\[0.5em]
	&\quad	
	+\,
	\frac{B[n(A - 2BX) - 2(2A - 3BX)] \phi ^{;\alpha }A_{;\alpha } \phi_{;\mu\nu}}{2 A (A - 2BX)^2} 
	\nonumber
	\\[0.5em]
	&\quad	
	+\,
	\frac{X B_{;\mu\nu}}{A - 2BX} 
	-
	\frac{[n(A - 2BX) - 2(A - 3BX)]  {A_{;\mu\nu}}}{2 A (A - 2BX)} 
	\nonumber
	\\[0.5em]
	&\quad	 
	-\, 
	\frac{B^{;\alpha} \phi_{;\alpha\nu} \phi_{;\mu}}{2 (A - 2BX)} 
	+ 
	\frac{B A^{;\alpha} \phi_{;\alpha\nu} \phi_{;\mu}}{2 A (A - 2BX)}
	+
	\frac{\phi^{;\alpha} B_{;\alpha\nu} \phi_{;\mu}}{2 (A - 2BX)}  
	\nonumber
	\\[0.5em]
	&\quad	
	-\,
	\frac{B \phi^{;\alpha} A_{;\alpha\nu} \phi_{;\mu}}{2 A (A - 2BX)}.
\end{align}
All terms on the right hand side of (\ref{hatRmnRmn}) are contributions from the disformal transformation involving the functional factors $A$ and $B$ and their derivatives. These derivatives go up to second order in $X$, which, in the language of $\phi$ translates to third order derivatives. This is in addition to the explicit lone term involving third orderivative of $\phi$; specifically, the one involving $\phi^{;\alpha}\phi_{;\mu\nu\alpha}$. As we shall see in the following section however, upon computing for the action and performing integration by parts, the derivative terms can go up to second order only for $\phi$. 

As a consistency check of the result above for the disformally transformed Ricci tensor, we can take its conformal limit to gain 
\begin{align}
    \label{hatRictens}
	\widehat R_{\mu \nu } - R_{\mu \nu }
	\,&\stackrel{\text{conf}}{=}\,
	\frac{3(n-2)A_{;\mu} A_{;\nu}}{4 A^2} 
	-
	\frac{(n-4)A^{;\alpha }A_{;\alpha }g_{\mu\nu}}{4 A^2} 
 	\nonumber
	\\[0.5em]
	&\qquad\quad	
	-\,
	\frac{(\square A) g_{\mu\nu}}{2 A}
	-
	\frac{(n-2){A_{;\mu\nu}}}{2 A},
\end{align}
This conformal limit is equivalent to the result in Ref. \cite{Wald:1984rg} (using a different notation).

\section{The transformed Einstein-Hilbert action}
\label{secdtehact}

\noindent 
Now that we have the transformed Ricci tensor, all we need to do is contract it with the inverse disformal metric and couple them with the spacetime integral measure to form the hatted action:
\begin{align}
    \widehat S
    &=
    \frac{1}{2}\int \dx^n\,\sqrt{-\widehat g}\,
    \widehat R.
\end{align}
The metric determinant above can be computed using the matrix determinant lemma. We then have
\begin{align}
	\widehat g
	&=
	gA^{n-1}(A - 2BX), 
\end{align}
which upon using the conformal-disformal constraint, can be rewritten simply as
\begin{align}
    \label{integmeas}
    \sqrt{-\widehat g}
    &=
    \sqrt{-g}\, bA^{n-1}
\end{align}
to form the spacetime integral measure, $\dd^n x \sqrt{-\widehat g} = \dx^n \sqrt{-g}\, bA^{n-1}$. Needless to say, the measure depends on the number of spacetime dimensions. This is in contrast to the case of constrained general disformal transformation in Ref. \cite{Alinea:2024gjn} where the coefficient of $\dd^n x$ only depends on the first power of the conformal factor.

On the other hand, the hatted Ricci scalar can be calculated as $\widehat R = \widehat g^{\mu\nu}\widehat R_{\mu\nu}$, with the inverse disformal metric and the hatted Ricci tensor given by (\ref{hatgmunu}) and (\ref{hatRmnRmn}), respectively. After a lengthy calculation---performing simplification and integration by parts---we find upon coupling $\sqrt{-\widehat g}$ above with $\widehat R$ the transformed action given by 
\begin{align}
    \label{hatSI}
	\widehat S
	&=
	\frac{1}{2}\int \dd^n x\sqrt{-g} \bigg\{
		bA^{n-2} R
	    -
        \frac{(n - 2)(n - 1)A_{,\phi }^2 X}{2bA^2}  
	   	\nonumber
	   	\\[0.5em]
	   	&\quad   
		+\,
		\frac{b^2A^{n-2} - 1}{2bX}
		[(\square \phi )^2 - \phi ^{;\mu \nu }\phi _{;\mu \nu }]	
	   	\nonumber
	   	\\[0.5em]
	   	&\quad  
	    -\,
	    \frac{(n - 2)(2b^2A^{n-2} - 1)}{bA}A_{,\phi}\, \square \phi       
	   	\nonumber
	   	\\[0.5em]
	   	&\quad
		+\,
	    \frac{n - 2}{2bA} \bigg[
	    	\frac{(n - 1)A_{,X}}{A}
			+
			\frac{2b^2A^{n-2} - 1}{X}	
	    \bigg]A_{,\phi}\phi ^{;\mu}X_{;\mu}   
		\nonumber
		\\[0.5em]
		&\quad      	 	 
		+\,
		\frac{(n - 2)A_{,X}}{4bAX} \bigg[
			\frac{(3n - 5)b^2A^{n-2} - (n - 1)}{2A} A_{,X}
    		\nonumber
    		\\[0.5em]
    		&\qquad\qquad\qquad\qquad    
			-\,
			\frac{b^2A^{n-2} - 1}{X}
	    \bigg]\phi ^{;\mu}X_{;\mu}\,\phi ^{;\nu}X_{;\nu}     	 \nonumber
	    \\[0.5em]
	    &\quad   
	    +\,
	    \bigg[
	    	bA^{n-3}\frac{n - 2}{2}\bigg(
	    		\frac{3 n - 5}{2A} A_{,X}^2
				+
				\frac{A_{,X}}{ X}
			\bigg)
    		\nonumber
    		\\[0.5em]
    		&\qquad\quad    
			-\,
			\frac{b^2A^{n-2} - 1}{2bX^2}     
		\bigg]X^{;\mu}X_{;\mu} 
	    \nonumber
	    \\[0.5em]
	    &\quad       
	    +\,
	    \frac{1}{2bX} \bigg[
		    (n - 2)\frac{2b^2A^{n-2} - 1}{A} A_{,X}
    	    \nonumber
    	    \\[0.5em]
    	    &\qquad\quad           
			-\,
			\frac{b^2A^{n-2} - 1}{X}
		\bigg]\phi ^{;\mu }X_{;\mu }\,\square \phi.
\end{align}

Note that the derivative of the conformal factor above is expanded as $A_{;\mu} = A_{,\phi} \phi_{;\mu} + A_{,X} X_{;\mu}$. Furthermore, a term involving the coupling of the derivative of scalar field and Ricci tensor, that is, $\phi^{;\mu}\phi^{;\nu}R_{\mu\nu}$ in $\widehat g^{\mu\nu}\widehat R_{\mu\nu}$, is re-expressed in terms of the (third)-order derivatives of scalar fields by virtue of the definition of the Riemann curvature tensor. Surprisingly, even with these expansions and the contraction of the lengthy equation for the hatted Ricci tensor and the bi-term inverse disformal metric, we have a remarkably short expression free from third-order derivatives of the scalar field. Such simplification is afforded by the use of the conformal-disformal constraint and good utility\footnote{Interestingly, this simple technique of integration can ``precipitate'' simple and physically meaningful results as can be seen for instance, in Refs. \cite{Maldacena:2002vr} and \cite{Alinea:2016glw}.} of the simple technique of integration by parts.

To make sense of the terms in the Lagrangian encoded in $\widehat S $ above, we map them to the Horndeski action \cite{Horndeski:1974wa}; that is, identify terms that can be matched with the sub-Lagrangians of this action. The Horndeski action describes the most general theory of gravity in four dimensions within the framework of scalar tensor theory, whose equations of motion are second order in nature. In other words, it describes a stable theory, free from Ostrogradsky instability and ghost terms \cite{Woodard:2015zca,Gleyzes:2014qga,Langlois:2015cwa,Kobayashi:2019hrl}. The action consists of four sub-Lagrangians $(\mathcal L_2, \mathcal L_3, \mathcal L_4, \mathcal L_5)$ and is given by 
\begin{align}
    S_H
    &=
    \int \sqrt{-g}\,\dd^4x\,[\mathcal L_2 + \mathcal L_3 + \mathcal L_4 + \mathcal L_5],
\end{align}
where
\begin{align}
    \mathcal L_2 
    &= 
    G_2,
    \nonumber
    \\[0.5em]
    \mathcal L_3 
    &= 
    G_3\,\square \phi,
    \nonumber
    \\[0.5em]
    \mathcal L_4 
    &= 
    G_4 R + G_{4,X}[(\square\phi)^2 - \phi_{;\mu\nu}\phi^{;\mu\nu}],
    \quad (G_{4,X} = \partial_X G_4)
    \nonumber
    \\[0.5em]
    \mathcal L_5
    &=
    -
    \frac{1}{3!}G_{5,X}[
        (\square\phi)^3 
        + 
        2\phi^{;\mu\alpha}\phi_{;\alpha\beta}{\phi^{;\beta}}_\mu
        -
        3\phi^{;\mu\nu}\phi_{;\mu\nu}\,\square\phi
    ]
    \nonumber
    \\[0.5em]
    &\quad
    +\,
    G_5 G_{\mu\nu}\phi^{;\mu\nu}, 
\end{align}
with $(G_2, G_3, G_4, G_5)$ being general functionals of $(\phi,X)$ and $G_{\mu\nu} \equiv R_{\mu\nu} - \frac{1}{2} g_{\mu\nu}R$, is the Einstein tensor.

Remarks are then in order with regards to the transformed Einstein-Hilbert action given by (\ref{hatSI}) in relation to the Horndeski action. First, with the equation for the Horndeski action in hand, in four dimensions, we can readily identify the second term in the integrand of $\widehat S$, being a functional of $(\phi,X)$, as belonging to $\mathcal L_2$; specifically,
\begin{align}
    -\frac{(n - 2)(n - 1)A_{,\phi }^2 X}{2bA^2} 
    \quad\subset\quad
    \mathcal L_2
\end{align}
Second, the term involving the d'Alembertian of $\phi$ belongs to the third sub-Lagrangian.
\begin{align}
    -
	\frac{(n - 2)(2b^2A^{n-2} - 1)}{bA}A_{,\phi} (\square \phi )
    \quad\subset\quad
    \mathcal L_3
\end{align}
Third, the inconspicuously written fourth term involving $\phi^{;\mu}X_{;\mu}$ ``belongs'', by fragmentation, to $\mathcal L_2$ and $\mathcal L_3$; see Refs. \cite{Alinea:2024gjn} and \cite{Bettoni:2013diz}. 

To see this, we let the functional coefficient of $\phi^{;\mu}X_{;\mu}$ in (\ref{hatSI}) be the $X$-derivative of the functional $H = H(\phi,X)$ defined as
\begin{align}
    \label{Hfunc}
    H
    &=
    \int^X\dd\bar X\,\frac{n - 2}{2bA} \bigg[
        \frac{(n - 1)A_{,X}}{A}
        +
        \frac{2b^2A^{n-2} - 1}{X}	
    \bigg]A_{,\phi}.
\end{align}
This allows us to rewrite the following term in (\ref{hatSI}) as
\begin{align}
    &\hspace{-1.0em}\int\dd^n x\sqrt{-g}\,
    \frac{n - 2}{2bA} \bigg[
        \frac{(n - 1)A_{,X}}{A}
        +
        \frac{2b^2A^{n-2} - 1}{X}	
    \bigg]A_{,\phi}\phi ^{;\mu}X_{;\mu} 
    \nonumber
    \\[0.5em]
    &
    =
    \int\dd^n x\sqrt{-g}\, (2H_{,\phi}X)
    -
    \int\dd^n x\sqrt{-g}\, H\,\square\phi,    
\end{align}
upon performing integration by parts. Clearly, the fourth term in the integrand in (\ref{hatSI}) belongs to sub-Lagrangians $\mathcal L_2$ and $\mathcal L_3$.

Fourth, we have the Ricci scalar in $\widehat S$ coupled with the conformal factor. In the Horndeski action, such a term belongs to $\mathcal L_4$. However, the accompanying term, $[(\square \phi )^2 - \phi ^{;\mu \nu }\phi _{;\mu \nu }]$, has a functional coefficient that does not, in general, correspond to the $X-$ derivative of the functional coefficient of $R$, needed to form $\mathcal L_4$. The sum of terms in $\widehat S$ given by 
\begin{align}   
    \label{beinl4}
    bA^{n-2} R
    +
    \frac{b^2A^{n-2} - 1}{2bX}
    [(\square \phi )^2 - \phi ^{;\mu \nu }\phi _{;\mu \nu }],
\end{align}
then begs for an auxiliary constraint on the $X$-dependence of $A$, namely,
\begin{align}
    \label{constrAX}
    (n-2)bA^{n -3} A_{,X}
    &=
    \frac{b^2A^{n-2} - 1}{2bX},
\end{align}
so that they can be made to belong $\mathcal L_4$. Such a constraint leads to the $X-$ functional dependence of the conformal factor being of the form $A^2 \sim \sqrt{|X|}$. In other words, the conformal factor retains its dependence on $(\phi,X)$ with the functional dependence on $\phi$ being the only free part.

Fifth and last, the last three terms in the integrand in (\ref{hatSI}), namely, those involving $(\phi^{;\mu}X_{;\mu})^2,\, X^{;\mu}X_{;\mu},$ and $\phi^{;\mu}X_{;\mu}\,\square \phi$ are beyond-Horndeski terms. Their corresponding equation of motion leads to differential equations beyond the second order. Ordinarily, this signals instability and ghost terms upon quantization. Nonetheless, there are healthy theories beyond the usual gamut of the Horndeski action subject to some invertibility conditions and additional constraints; see Refs. \cite{Gleyzes:2014qga,Langlois:2015cwa,Kobayashi:2019hrl}. For the three beyond-Horndeski terms in (\ref{hatSI}), owing to their complexity and the accompanying challenges with taming instabilities, we find it prudent to leave them for future work. 

With all the terms accounted for in $\widehat S$ above and the constraints put in place, we can rewrite the full action as
\begin{align}
    \label{hatSII}
	\widehat S
	&=
	\frac{1}{2}\int \dd^n x\sqrt{-g} \bigg\{
		2H_{,\phi}X 
	    -
        \frac{(n - 2)(n - 1)A_{,\phi }^2 X}{2bA^2} 
	   	\nonumber
	   	\\[0.5em]
	   	&\quad   
        -\,
        \bigg[
            H
            +
            \frac{(n - 2)(2b^2A^{n-2} - 1)}{bA}A_{,\phi}
        \bigg]\square \phi
	   	\nonumber
	   	\\[0.5em]
	   	&\quad   
		+\,
        bA^{n-2} R
        +
		(n-2)bA^{n-3}A_{,X}
		[(\square \phi )^2 - \phi ^{;\mu \nu }\phi _{;\mu \nu }]
		\nonumber
		\\[0.5em]
		&\quad      	 	 
		-\,
		\frac{(n - 2)A_{,X}^2}{8bA^2X} [
            (n-3)b^2A^{n-2} + (n - 1)
	    ]\phi ^{;\mu}X_{;\mu}\,\phi ^{;\nu}X_{;\nu}     	 \nonumber
	    \\[0.5em]
	    &\quad   
	    +\,
	    \frac{(n-2)bA^{n -3}A_{,X}}{2}        
        \bigg(
            \frac{3 n - 5}{2A} A_{,X}
            -
            \frac{1}{ X}
        \bigg)  
		X^{;\mu}X_{;\mu} 
	    \nonumber
	    \\[0.5em]
	    &\quad       
	    -\,
	    \frac{(n - 2)A_{,X}}{2bAX}
        \phi ^{;\mu }X_{;\mu }\,\square \phi\bigg\},
\end{align}
or, specifically in four dimensions we have upon further setting\footnote{One may wonder if the parameter $b$ can be adjusted to some other values so as to further simplify the resulting action. However, a short inspection of the equation for $\hat S$ above tells us that $b$ always appear ``coupled'' with the conformal factor; e.g. $2b^2A^{n-2}-1$. It follows that no amount of adjustment of $b$ apart from its trivial value can eliminate additional terms in the action beyond those existent with $b=1$. Furthermore, one may think of promoting $b$ to a functional of $\phi$ or $(\phi, X)$ and perhaps possibly consider a more general constraint stemming from the action with $X^p$ instead of simply $X$ for the KG action. But then again, based on our calculations, these only lead to a more complicated result in $n$ dimensions.} $b=1$,
\begin{align}
    \label{hatSI4}
	\widehat S^{(4)}
	&=
	\int \dd^4 x\sqrt{-g} \bigg\{
		2H_{,\phi}X
	    -
        \frac{3A_{,\phi }^2 X}{A^2}    
	   	\nonumber
	   	\\[0.5em]
	   	&\quad
        -\,
        \bigg[
            H
            +
	       \frac{2(2A^{2} - 1)}{A}A_{,\phi}
        \bigg]\square \phi
	   	\nonumber
	   	\\[0.5em]
	   	&\quad        
        +\,
        A^2 R
		+
		2AA_{,X}
		[(\square \phi )^2 - \phi ^{;\mu \nu }\phi _{;\mu \nu }]
		\nonumber
		\\[0.5em]
		&\quad      	 	 
		-\,
		\frac{A_{,X}^2}{4A^2X}(A^2 + 3)
        \phi ^{;\mu}X_{;\mu}\,\phi ^{;\nu}X_{;\nu}     	 
	    \\[0.5em]
	    &\quad   
	    -\,
	    \frac{A_{,X}}{8AX}(A^2 + 7)X^{;\mu}X_{;\mu}      
	    -
	    \frac{A_{,X}}{AX} \phi ^{;\mu }X_{;\mu }\,\square \phi \bigg\}.
        \nonumber
\end{align}

The KG-constrained Bekenstein disformal transformation, thus, leads to Horndeski action sans the fifth sub-Lagrangian and with three additional beyond-Horndeski terms. Comparing this with the \textit{special} disformal transformation of the Einstein-Hilbert action leading to a stable action consisting of terms under $\mathcal L_2,\, \mathcal L_3,$ and $\mathcal L_4$ alone \cite{Alinea:2020sei}, it is clear that the `extraneous' beyond-Horndeski terms are brought about by the $X$-dependence of the conformal and disformal factors. Having said this, the number and length of extraneous terms could have been far more complicated than what we have above. We have significantly ``tamed'' the resulting action coming from the lengthy contraction of $\sqrt{-\widehat g}\,\widehat g^{\mu\nu}\widehat R_{\mu\nu}$. It is with the conformal-disformal constraint rooted in the invariance of the massless Klein-Gordon equation under the Bekenstein transformation that we are able to afford a concise transformed action given by (\ref{hatSI4}). 

From another perspective, we are fortunate enough to have found a relatively simple transformed action given that we have started with a lone term---the Ricci scalar---in the Lagrangian. In a related study in Ref. \cite{Bettoni:2013diz}, the authors started with the full Horndeski action for the \textit{special} disformal transformation. Upon transformation, the resulting hierarchical propagation of terms coming from multiple sub-Lagrangians effectively lead to cancellation of ``extraneous'' terms. While ours is certainly spoiled by the additional dependence on the kinetic term in the Bekenstein disformal transformation \cite{Alinea:2020sei}, we have no other terms in the original action to begin with, to possibly cancel ``extraneous'' terms in the resulting transformed action. For future work, it would be interesting to modify the Einstein-Hilbert action by adding some sort of ``counter terms'' to cancel ``extraneous'' terms following the modified action's disformal transformation.

\textbf{Case of $\mathbf{n = 3}$ and $\mathbf{n = 2}$ dimensions.} It is interesting to know the structure of the transformed action in higher and lower dimensions\footnote{Interested readers may see Ref. \cite{Kobayashi:2019hrl} for a review of Horndeski theory and beyond, and some insights in and value of higher and lower dimensional theories. In addition to this, the cases for $n = 3$ and $n =2$ for constrained general disformal transformation are also covered in Ref. \cite{Alinea:2024gjn}.} compared to $n = 4$. Looking at the form of $\widehat S$ given by (\ref{hatSII}), however, cases for which $n > 4$ seem to yield a more complicated result. On the other hand, the presence of powers of $A^{n-3}$ and factors of $n-2$ in (\ref{hatSII}) are suggestive of simpler actions for $n < 4$, so we focus on cases for lower dimensions. 

For $n = 3$ and $b = 1$ we find the hatted action given by
\begin{align}
	\widehat S^{(3)}
	&=
	\int \dd^3 x\sqrt{-g} \bigg\{
		2XH_{,\phi }
		-
	    \frac{A_{\phi }^2 X}{A^2}		
	   	\nonumber
	   	\\[0.5em]
	   	&\quad		
	   	-\,
	   	\bigg(
		   	H 
		    +
		    \frac{2A - 1}{A}A_\phi
		\bigg)\square \phi     
		+
		A R
	    \nonumber
	    \\[0.5em]
	    &\quad   
		+\,
		A_X
		[(\square \phi )^2 - \phi ^{;\mu \nu }\phi _{;\mu \nu }]
		-
		\frac{A_{X}^2}{4A^{3}X} \phi ^{;\mu}X_{;\mu}\,\phi ^{;\nu}X_{;\nu}     	 	     	
	    \nonumber
	    \\[0.5em]
	    &\quad   
	    -\,
	    \frac{A_{X}}{2AX}X^{;\mu}X_{;\mu} 
	    -
	    \frac{A_{X}}{2AX} 
	    \phi ^{;\mu }X_{;\mu }\,\square \phi
	\bigg\},
\end{align}
where
\begin{align}
	H
	&=
	\int^X\dd\bar X\frac{2A^2 - 1}{2A^2\bar X}
	A_\phi
\end{align}
Observe that we do not have a reduction in the overall number of terms in the corresponding Lagrangian. The first five terms in the Lagrangian, similar to the case for $n = 4$, are the Horndeski-like terms\footnote{The original Horndeski action is specific for $n = 4$ dimensions so we simply call here terms analogous to the terms in the Horndeski action, ``Horndeski-like'' terms.}. On the other hand, the three ``extraneous'' terms involving $(\phi^{;\mu}X_{;\mu})^2,\, X^{;\mu}X_{;\mu},$ and $\phi^{;\mu}X_{;\mu}\,\square \phi$ remain in the action. What has changed are the functional coefficients of these terms; e.g., for $X^{;\mu}X_{;\mu}$, we have $(7 + A^2)A_{,X}/8AX$ to $A_{,X}/2AX$ from $n = 4$ to $n = 3$ dimensions. There is a marked simplification in these functional coefficients from $n = 4$ to $n = 3$ dimensions. For the ``extraneous'' terms, the functional coefficients involve $A_{,X}$ and the kinetic term $X$ explicitly, and this holds even in the absence of the auxiliary constraint (\ref{constrAX}). This effectively marks their existence being substantially rooted in the $X$-dependence of the disformal transformation even under the conformal-disformal constraint. Upon lifting the auxiliary constraint\footnote{We should be careful with simply letting $A_{,X} = 0$ with the auxiliary constraint (\ref{constrAX}) turned on. In such a scenario, the disformal factor vanishes and the conformal factor becomes (trivially) a constant.} (\ref{constrAX}), when $A_X = 0$, the ``extraneous'' term involving $(\phi^{;\mu}X_{;\mu})^2$ vanishes while the other two remains with equal functional coefficients of $(A^2 - 1)/2X^2$ in the Lagrangian corresponding to $S^{(3)}$.

For $n = 2$ and $b = 1$ we find from (\ref{hatSII}) the transformed action given by 
\begin{align}
    \label{hatSIII}
	\widehat S^{(2)}
	&=
	\frac{1}{2}\int \dd^n x\sqrt{-g} (
		2H_{,\phi}X 
        -
        H\square \phi
        +
        R
    ).
\end{align}
All the `extraneous' vanish and we are only left with the Ricci scalar and terms involving $H$. In fact, the latter vanishes by virtue of $(\ref{Hfunc})$. We are then left with 
\begin{align}
    \widehat S^{(2)}
    &=
    S^{(2)}.
\end{align}
In other words, the Einstein-Hilbert action is invariant under the constrained Bekenstein disformal transformation in $n=2$ dimensions! 

To further examine this, we trace from the conformal-disformal constraint given by (\ref{confdisfcon}) that the disformal factor vanishes in $n = 2$ dimensions. In other words, the disformal transformation becomes a conformal transformation. It follows from (\ref{hatRictens}) that the disformal Ricci tensor becomes related to the original one as
\begin{align}
	\widehat R^{(2)}_{\mu \nu } - R^{(2)}_{\mu \nu }
	&=
	\frac{A^{;\alpha }A_{;\alpha }g_{\mu\nu}}{2A^2} 
	-
	\frac{(\square A) g_{\mu\nu}}{2 A}.
\end{align}
Observe that the Ricci tensor itself is not invariant under this (reduced) conformal transformation. Nonetheless, the integral measure in two dimensions becomes simply $\dd^2 x\sqrt{-g}A$ from (\ref{integmeas}) while the inverse disformal metric turns to $\widehat g^{\mu\nu} = g^{\mu\nu}/A$ by virtue of (\ref{hatgmunu}). In effect, in two dimensions, the difference between the hatted Ricci scalar and the original one is given by
\begin{align}
    \widehat R^{(2)} - R^{(2)}
    &=
    \frac{A^{;\alpha}A_{;\alpha}}{A^2}
    -
    \frac{\square A}{A}.
\end{align}
The two terms on the right hand cancel upon integration by parts. Thus, we reach the same conclusion above. The Einstein Hilbert action is invariant under the Bekenstein's disformal transformation subject to the conformal-disformal constraint that, likewise, makes the massless Klein-Gordon equation invariant.

In hindsight, the fact that the Einstein-Hilbert action is invariant (up to a boundary term)\footnote{We are grateful to the kind reviewer of this paper for the lucid explanation of the general invariance (up to a surface term,) of the Einstein-Hilbert action in two dimensions.} under the Bekenstein's disformal transformation in $ n = 2 $ dimensions, should not be surprising at all; that is, even in the absence of the conformal-disformal constraint. To see this, we note that in general, the quantity $ \sqrt{-g}R  $ in the Einstein-Hilbert action can be written as \cite{Jensko:2023}
\begin{align}
	\sqrt{-g} R 
	=&
	\sqrt{-g}\, g^{\mu \nu }(
		\Gamma ^\lambda _{\mu \sigma }\Gamma ^\sigma _{\lambda \nu }
		-
		\Gamma ^\sigma _{\mu \nu }\Gamma ^\lambda _{\lambda \sigma }
	)
	\nonumber\\ &+
	\partial _\sigma [\sqrt{-g}(
		g^{\mu \nu }\Gamma ^\sigma _{\mu \nu }
		-
		g^{\sigma \nu }\Gamma ^\lambda _{\lambda \nu }
	) ].
\end{align}
In other words, the action can be decomposed as a bulk term from which the Einstein field equations can be derived, and a surface or boundary term. In $ n = 2 $ dimensions, the action is simply a boundary term; it becomes topological \footnote{The fact that the Einstein-Hilbert action is topological in nature in $ n = 2 $ dimensions, poses some problems in quantum gravity due to the suppression of fluctuations. To address this, one may include a dilaton field coupled with the Ricci scalar in the Einstein Hilbert action as in the so-called Jackiw-Teitelboim (JT) action \cite{Jackiw:1984je,Teitelboim:1983ux}. The JT action can be recast into a Kinetic Gravity Braiding (KGB) theory---a 2D Horndeski theory. The latter in turn, is closed under Bekenstein's disformal transformation \cite{Takahashi:2018yzc,Nejati:2023hpe}.} in nature. On performing conformal or disformal transformation, the total derivative in the action remains\footnote{Note that it can be challenging to show this following a brute-force calculation of the Bekenstein's disformal transformation of $ \sqrt{-g}R $. This is because in $ n = 2 $ dimensions, $ \sqrt{-\widehat g}  = \sqrt{-g}\sqrt{A(A- 2BX)}$ while $ \widehat R $ only involves rational functions of $(A,B,X)$ and derivatives thereof. Bypassing this challenge, in some sense, the Klein-Gordon constraint, `folds' $ \widehat S^{(2)} $ back to $ S^{(2)} $ as it should be, respecting the topological nature of the Einstein-Hilbert action in two dimensions.} just that, a total derivative. Hence, the action remains a boundary term and $ S^{(2)} = \widehat S^{(2)} $.

\section{Summary and concluding remarks}
\label{seConclude}
The General Theory of Relativity is the framework for Modern Cosmology, the basis of our current understanding of spacetime, and the justification for numerous astrophysical phenomena. At its core is the Einstein-Hilbert action that encodes its beautiful mathematical structure. In spite of its beauty and far-reaching applications, the General Theory of Relativity is far from a complete theory. In addition to this, we believe that there are still a lot of avenues to investigate its nature with regards to its symmetry (or asymmetry) and possible extension beyond the current realm that it dominates.

Motivated by this belief, we explore in this study the Bekenstein's disformal transformation of the Einstein-Hilbert action. This is in the hope that we may be able to connect or extend it to the Horndeski action and to some extent, gain further insight about its (symmetry) tranformation properties. However, in our attempt to perform transformation, the result in sight is an unwieldy action; extremely long and complicated expression involving the conformal and disformal functional factors, and their derivatives. To `tame' the transformed action we exploit the conformal-disformal constraint discovered in a previous study on the invariance of the massless Klein-Gordon equation. 

Upon application of this constraint, we find a remarkably more concise transformed action in four dimensions. The corresponding sub-Lagrangians may be mapped to three out of four sub-Lagrangians of the Horndeski action in Scalar Tensor theory, with three additional beyond-Horndeski terms. The latter group of three terms may be rooted in the kinetic term dependence of the Bekenstein's disformal transformation. This explanation stems from a related study on the \textit{special} disformal transformation of the Einstein-Hilbert action, wherein the resulting action has its Lagrangian that can be completely mapped to three out of four sub-Lagrangians of the Horndeski action; that is, in the absence of kinetic term-dependence in the metric transformation, there exists no `extraneous' terms. Unfortunately, the signature of these terms are carried to the case of three spacetime dimensions. Only in two dimensions, do these terms non-trivially vanish. The Einstein-Hilbert action in two dimensions is invariant under the KG-constrained Bekenstein's disformal transformation. In fact, even without this constraint, the action should be invariant (up to a boundary term) because of its topological nature in two dimensions.

For our future work, we would like to further investigate the beyond-Horndeski terms in the transformed Einstein-Hilbert action, their deep mathematical meanings and physical implications beyond what the current literature offers. Our insight is that the first two shall complement each other in relation to possible supplementary constraints, group theoretic property of disformal transformation, and unbounded energy in the corresponding system that they could describe. Beyond this, we would like to also explore the possible complete `removal' of beyond-Horndeski terms, at least in four spacetime dimensions, from the resulting action upon Bekenstein's disformal transformation {(or some other forms of the disformal transformation)} of the Einstein-Hilbert action { and possibly, its closely related variations}. 


\newpage
\end{multicols}
\end{document}